\documentclass[12pt]{iopart}

\usepackage{iopams}
\usepackage{setstack}
\usepackage{graphicx}
\DeclareGraphicsExtensions{.eps,.ps}
\usepackage{harvard}
\usepackage{nicefrac}
\usepackage{color}
\usepackage{booktabs}
\usepackage{textcomp}

\begin{document}

\title{Response of a ferrofluid to traveling-stripe forcing}

\author{A Beetz, C Gollwitzer, R Richter and I Rehberg}

\address{Experimentalphysik V
Universit\"atsstr. 30 95445 Bayreuth, Germany}
\ead{Christian.Gollwitzer@uni-bayreuth.de}

\begin{abstract}
We observe the dynamics of waves propagating on the surface of a ferrofluid under the influence of a spatially and temporarily modulated field. In particular, we excite plane waves by a travelling lamellar modulation of the magnetization. By this external driving both the wavelength and the propagation velocity of the waves can be controlled. The amplitude of the excited waves exhibits a resonance phenomenon similar to that of a forced harmonic oscillator. Its analysis reveals the dispersion relation of the free surface waves, from which the critical magnetic field for the onset of the Rosensweig instability can be extrapolated.
\end{abstract}

\maketitle

\section{Introduction}
When a critical value of the vertical magnetic induction is surpassed, the surface of a ferrofluid exhibits an array of liquid crests. This so called Rosensweig instability \cite{cowley1967} has been investigated in a static field \citeaffixed{bacri1984,richter2005,gollwitzer2007}{see e.\,g.} and under temporal modulation of either the magnetic field \citeaffixed{mahr1998c}{see e.\,g.} or the gravitational acceleration \cite{ko2003}. For a recent survey see \citeasnoun{richter2007}. While these excitations are homogeneous in space, a combined spatiotemporal forcing of the plain surface has been 
implemented by \citeasnoun{kikura1990} using an array of solenoids. They investigate the surface waves for small magnetic fields far below the Rosensweig threshold  and measure the resulting volume flow rate. In contrast, we apply a traveling-stripe forcing to the surface of magnetic liquids in the advent of the Rosensweig instability and uncover a resonance phenomenon of the wave amplitude with respect to the lateral driving velocity. So far, such  
a spatiotemporal driving of a \emph{pattern forming system} has only been realized in a chemical experiment by \citeasnoun{miguez2003}. A review by \citeasnoun{ruediger2007} calls for further experiments, and the present paper provides one.

\begin{figure}[b!]
  \centering
    \resizebox{.65\linewidth}{!}{\input{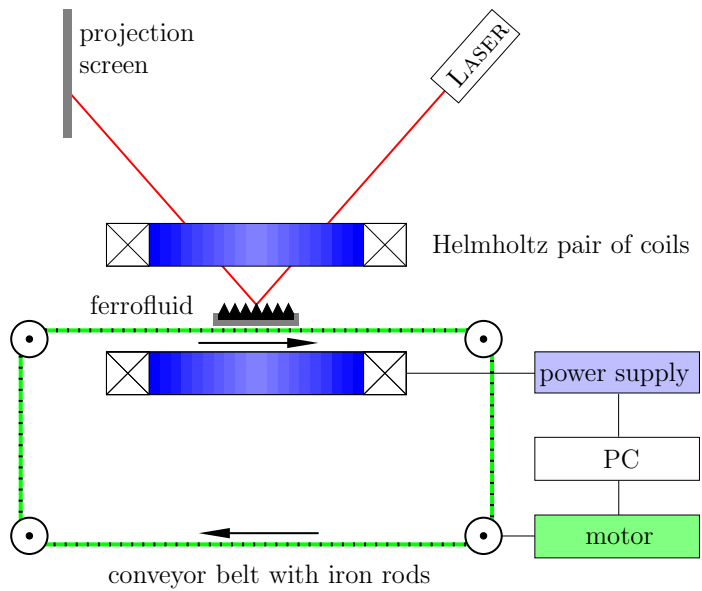}}
    \caption{Sketch of the experimental setup} \label{Exp-Setup}
\end{figure}

\section{Experimental}
\begin{figure}[htb!]
  \centering
    (a)\raisebox{0.05\linewidth}{\includegraphics[height=.45\linewidth,angle=0]{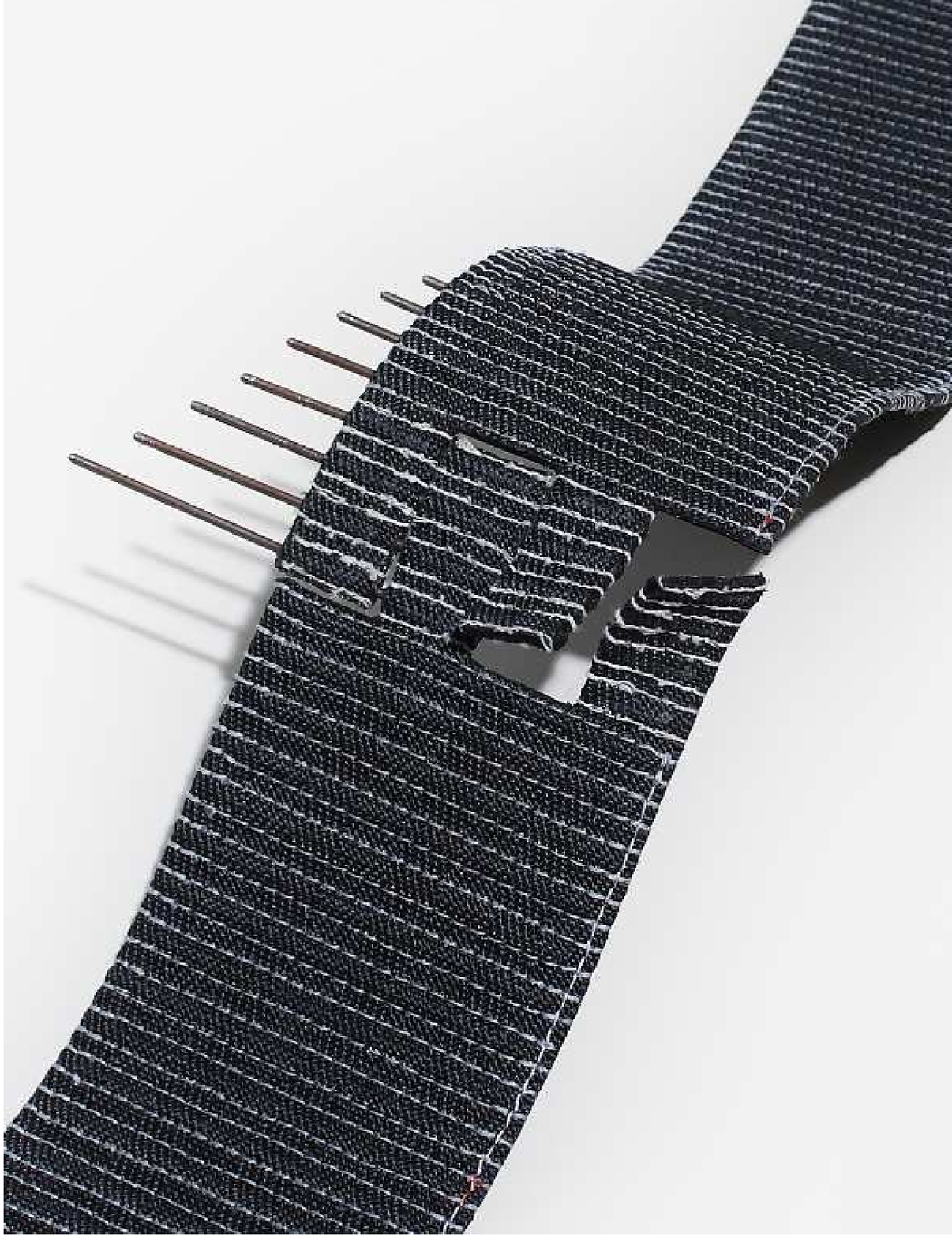}}
    (b)\raisebox{0.05\linewidth}{\includegraphics[height=.45\linewidth]{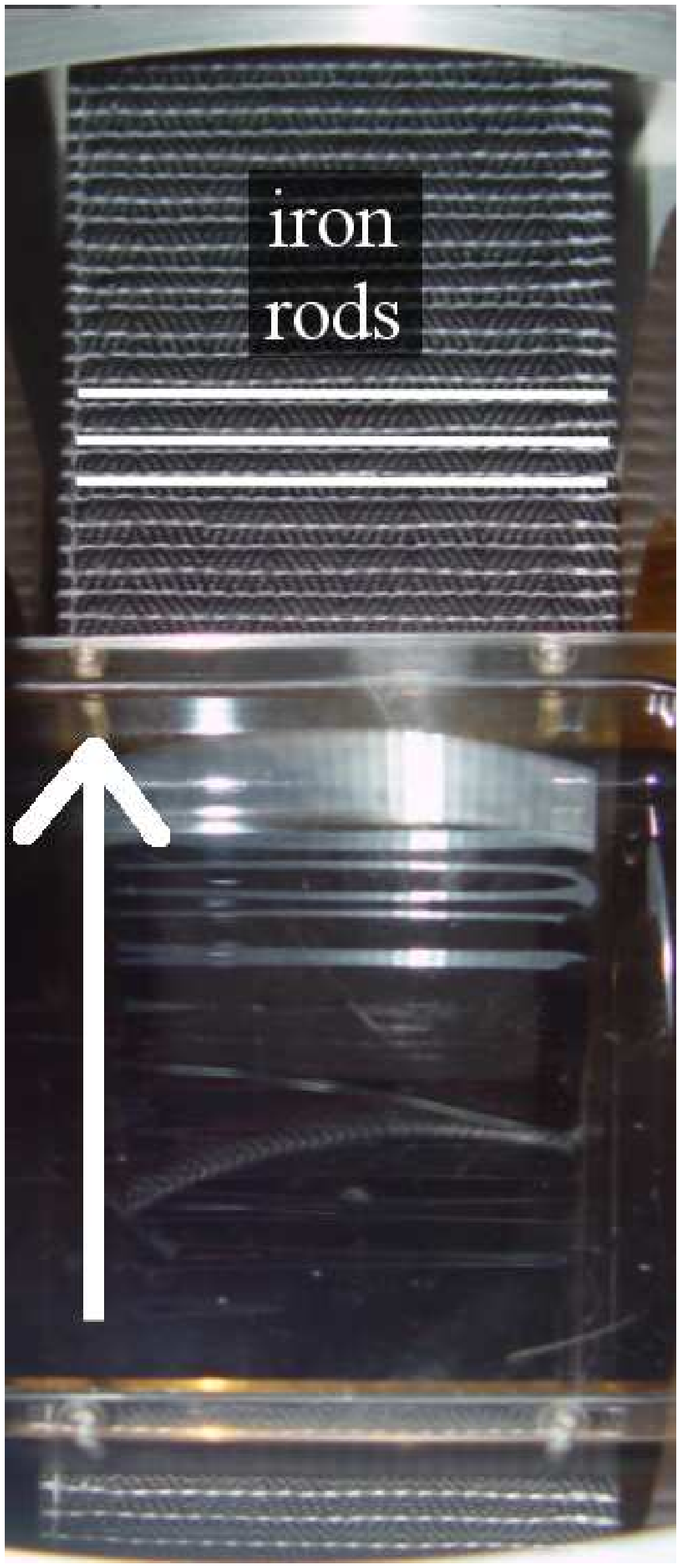}}
    \caption{Experimental realization of the spatiotemporal excitation:
      Conveyor belt with the iron rods (a), and view on the ferrofluid container from
      above (b). Note the surface undulations in the upper part of the container due
      to traveling-stripe forcing.} \label{conveyorbelt}
\end{figure}

\begin{figure}[htb!]
  \centering 
    \input{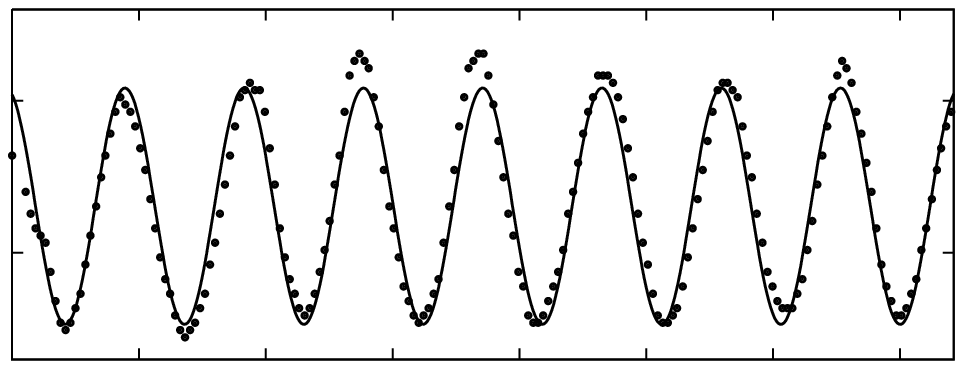} 
    \caption{Modulation of the vertical
      magnetic field along the conveyor belt (circles), and its sinusoidal fit (solid
      line) by $B(x)= B_0+ \Delta B \sin(k_c x)$, with $B_0=15.15\,\mathrm{mT}$ and
      $\Delta B=(0.39\pm0.01)\,\mathrm{mT}$.} \label{magnetmod}
\end{figure}

Our experimental setup is sketched in figure~\ref{Exp-Setup}. A rectangular vessel machined from Perspex\texttrademark \ is placed in the center. The inner dimensions are 100\,mm\,(x), 120\,mm\,(y), 25\,mm\,(z). It is filled with ferrofluid up to $z=3\,$mm. A  Helmholtz pair of coils generates a constant magnetic induction which is homogenous with a deviation of 1\,\% over the size of the container. In addition we apply a small (5\,\%) spatiotemporal modulation of the magnetic field with the spatial periodicity of the critical wavelength $\lambda_\mathrm{c}=2\pi\sqrt{\sigma/(\rho g)}$, which propagates parallel to the surface with constant velocity $v$. Then traveling waves at the surface of the fluid are formed with the same wavelength and speed as that of the driving. The field modulation is realized by a ``conveyor belt'', made from a textile band which harbors periodically placed iron rods. The lateral distance between neighbouring rods was selected to be as close as possible to the critical wavelength $\lambda_\mathrm{c}=9.98\,\mathrm{mm}$. We achieved $9.3\pm1\,\mathrm{mm}$. The rods have a length of $70\,\mathrm{mm}$ and are made of welding wire with a diameter of $1.0\,\mathrm{mm}$. The vertical distance between the symmetry axis of the rods and the surface of the fluid at rest is $8\pm0.5\,\mathrm{mm}$. The band is driven by an electric motor which allows to vary the velocity up to $30\,\mathrm{cm/s}$. Figure~\ref{conveyorbelt}\,(a) shows the bare belt and (b) its location and its effect within the setup. The iron rods amplify the magnetic field locally due to their higher susceptibility, thus the magnetic field strength varies along the driving direction. As demonstrated in figure~\ref{magnetmod}, the excitation profile is approximately sinusoidal. We investigate the response of the ferrofluids APG\,512a and EMG\,909 (Ferrotec Co.). The parameters of those fluids are listed in table~\ref{ferrodaten}.
\begin{table}[h!]
  \centering
  \caption{Parameters for the ferrofluids APG\,512a Lot~083094CX and EMG\,909 Lot~F050903B. $B_\mathrm{c}$ is computed from the material parameters, while $B_\mathrm{c,m}$ corresponds to the results obtained in this paper.}
  \label{ferrodaten}
  \begin{tabular}{llp{0.4cm}r@{}lp{0.4cm}r@{}l}
\toprule
property        &          && \multicolumn{2}{c}{APG\,512a}      && \multicolumn{2}{c}{EMG\,909} \\ \cmidrule{1-8}
density         &  $\rho$  && $1.26\ $&$\mathrm{g/cm}^3$   &&  $1.0047\ $&$\mathrm{g/cm}^3$    \\
surface tension & $\sigma$ && $30.57\ $&$\mathrm{mN/m}$    &&  $24.51\ $&$\mathrm{mN/m}$       \\
initial susceptibility & $\chi_0$ && $1.172\ $ &           &&  $0.760\ $   &           \\
viscosity       & $\eta$   && $120\ $&$\mathrm{mPa\,s}$    &&  $4.7\ $&$\mathrm{mPa\,s}$       \\
critical wavelength & $\lambda_\mathrm{c}$ && $9.98\ $&$\mathrm{mm}$  &&  $9.81\ $&$\mathrm{mm}$  \\
computed critical induction & $B_\mathrm{c}$ && $17.22\ $&$\mathrm{mT}$ &&
$25.95\ $&$\mathrm{mT}$\\ \cmidrule{1-8}
measured critical induction & $B_\mathrm{c,m}$  && $15.88\ $&$\mathrm{mT}$  &&  $23.64\ $&$\mathrm{mT}$ \\
\bottomrule
  \end{tabular}
\end{table}

\setlength{\unitlength}{\linewidth}

For measuring the amplitudes we direct a beam of a helium-neon laser onto the surface of the magnetic fluid in the middle of the container, from where it is reflected to a screen. The beam position on the screen is acquired via a charge-coupled-device (CCD) camera and reveals the slope of the surface at the incident point of the laser beam. We extract the height of the undulations $A$ by assuming that the surface modulation is sinusoidal, which is valid in this case of only small deformations. When the conveyor belt moves under the vessel, the fluid ridges travel with the iron rods and the position of the reflected beam oscillates.

Similar to figure~\ref{magnetmod}, the surface undulations are approximately sinusoidal. Deviations stem from inaccuracies in the spacing of the iron rod lattice, and fluctuations in the driving velocity of about 0.1\,\%, which results in a phase jitter of $\pm\,\pi/2$. The noise spectrum extends above and below the modulation frequency, and can be removed by applying a bandpass filter centered around the mean frequency of the moving iron rods.

The amplitude of the surface waves is determined from the fourier spectrum by adding the intensities of the mode corresponding to the mean passage time of one rod and the two neighbouring modes. The spectrum is calculated within a passage time of 314 rods, i.e. one complete cycle of the conveyor belt. Because the data can be continued periodically, a window function for the elimination of artifacts is not needed. 

\begin{figure}
  \centering 
    (a) \raisebox{0.05\linewidth}{\input{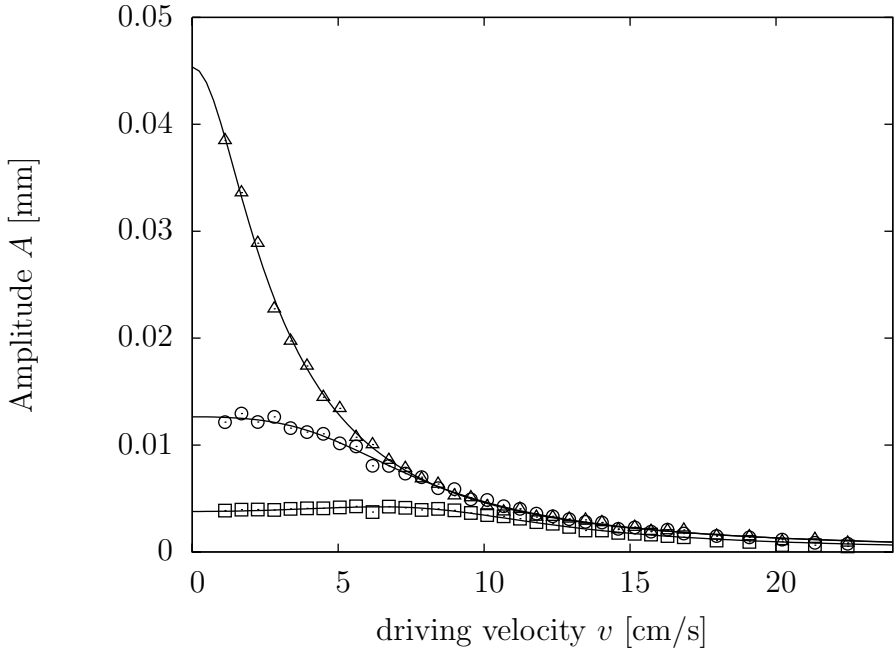}}\\
    (b) \raisebox{0.05\linewidth}{\input{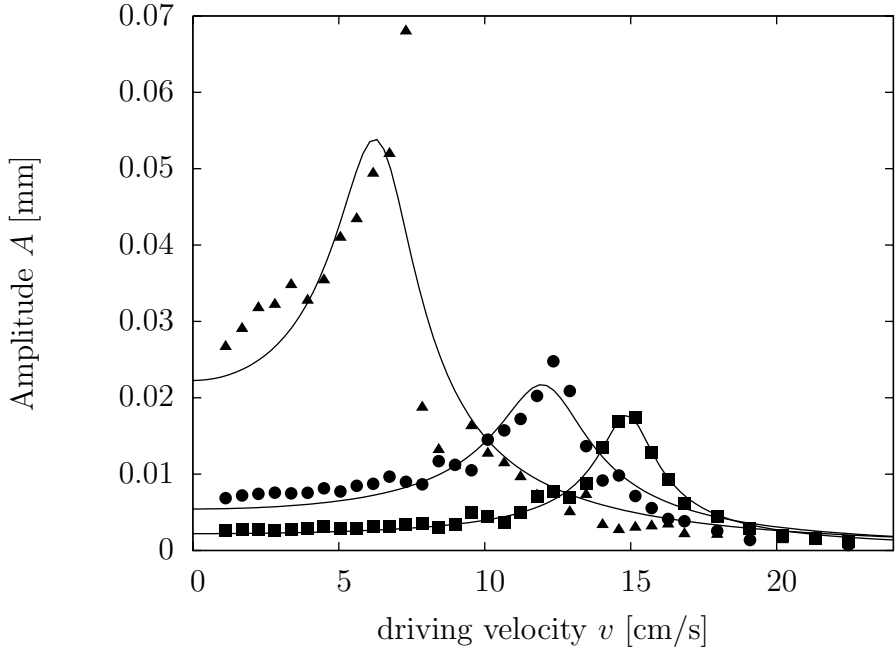}}
    \caption[] {Amplitude of the
      waves versus the driving velocity for varying magnetic induction for 
      the ferrofluids APG\,512a (a) and EMG\,909 (b). The triangles mark 
      the highest induction, the circles an interim value and the boxes 
      the lowest field. The data have been captured at following magnetic inductions: 
      $\bigtriangleup$:~$15.0\,\mathrm{mT}$, $\odot$:~$13.4\,\mathrm{mT}$, 
      $\boxdot$:~$10.4\,\mathrm{mT}$, $\blacktriangle$:~$20.9\,\mathrm{mT}$, 
      \fullcircle:~$14.9\,\mathrm{mT}$, $\blacksquare$:~$10.4\,\mathrm{mT}$. 
      The solid lines display a fit by (\ref{amplitudeeq}).} \label{amplitude}
\end{figure}

\section{Experimental results}
We have measured the amplitude of the fluid waves below the critical field of the Rosensweig instability as a function of the driving velocity and the applied magnetic field for two different magnetic fluids. For each fluid we selected about 10 different magnetic inductions in a range where visible undulations occur. The results for three representative values of $B$ are displayed in figure~\ref{amplitude}. For slow driving velocities the amplitude of the undulations resembles that of the static case. For increasing velocity, the amplitude of the traveling waves passes through a maximum and decays for high driving velocities. When increasing the magnetic field the undulations become remarkably higher and the maximum shifts to lower driving speeds. When the critical field is reached, the undulations are replaced by  Rosensweig spikes. Note the different response of the highly viscous (a) and the less viscous fluid (b).

Below the Rosensweig threshold $B_c$, the amplitude response to this spatiotemporal driving can be modeled as a damped forced harmonic oscillator
\begin{equation}
A(v)=\frac{A_0 v_p^2}{\sqrt{({v_p^2}-v^2)^2+4\gamma^2 v^2}}. \label{amplitudeeq}
\end{equation}
Here $A(v)$ denotes the amplitude dependent on the driving velocity $v$, where $A_0$ is the amplitude at zero velocity, $v_p$ denotes the phase velocity of the unforced surface waves and $\gamma$ the damping constant. As displayed in figure~\ref{amplitude}, the experimental data are well captured by fits to (\ref{amplitudeeq}).

The viscosity determines the damping constant in our model (\ref{amplitudeeq}). This is corroborated by figure~\ref{amplitude}\,(a), where the curves of the highly viscous APG\,512a show an over-damped behavior. In contrast the fluid EMG\,909, with a 25 times smaller viscosity, displays a clearly visible maximum for all magnetic inductions (see b).

The resonant propagation speed $v_p$ can be calculated from the dispersion relation. For infinite layer thickness and inviscid fluids the surface waves on ferrofluid are described by the plain dispersion relation
\begin{equation}
\omega^2=gk+\frac{\sigma}{\rho}k^3-\frac{1}{\rho}\frac{(\mu_\mathrm{r}-1)^2}{
\mu_0\mu_\mathrm{r}(\mu_\mathrm{r}+1)}B^2k^2 \label{dispersion}
\end{equation}
put forward by \citeasnoun{cowley1967}. In our experiment we intentionally constrain the wavenumber of the surface waves by the iron rods to $k_c$. From (\ref{dispersion}), we get a phase velocity $v_p=\frac{\omega}{k_c}$ which depends on the magnetic induction:
\begin{equation}
\fl \qquad
v_p = \sqrt{\frac{2}{\rho}\sqrt{g \rho \sigma}-\frac{1}{\rho}\frac{(\mu_\mathrm{r}-1)^2}{\mu_0\mu_\mathrm{r} (\mu_\mathrm{r}+1)} B^2} 
    = \sqrt{\alpha \ \frac{B_c^2-B^2}{B_c^2} }\ ,\qquad \label{wurzelfit}
\mathrm{with}\ \alpha = 2 \sqrt{\frac{g \sigma}{\rho}}\ .
\end{equation}
\begin{figure}[htb]
  \centering
    \input{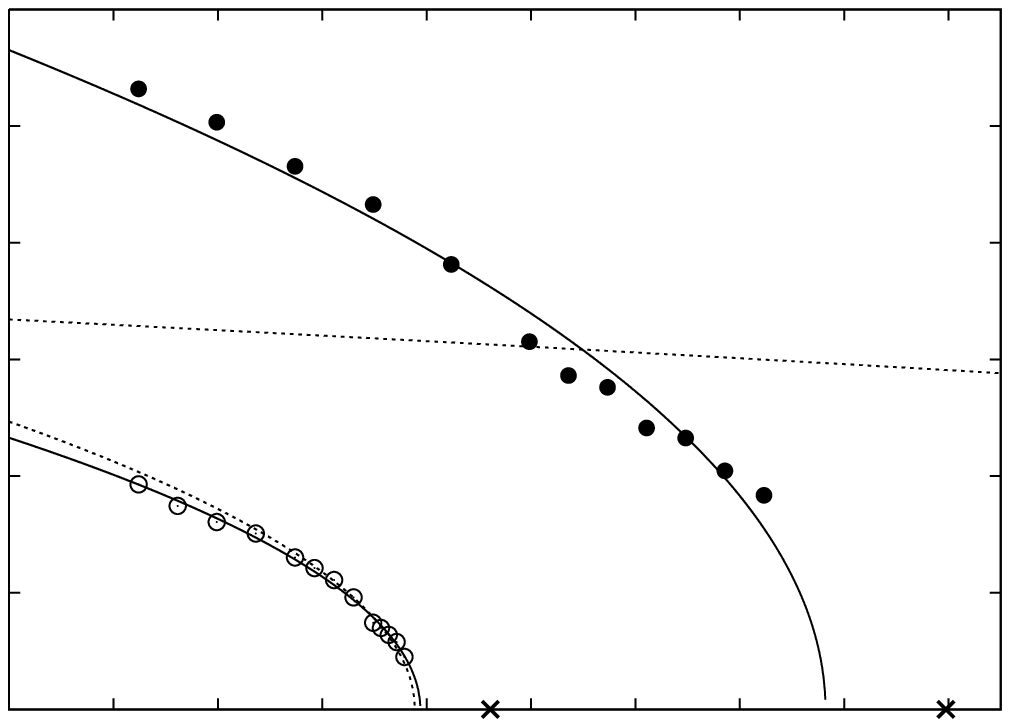}
    \caption{The resonant velocity for APG\,512a (empty circles) and EMG\,909
      (full circles) extracted from the fit by (\ref{amplitudeeq}) plotted against 
      the magnetic induction. The curves display fits by (\ref{wurzelfit}). 
      The dashed curve results from a one parameter fit taking into account the
      material parameters whereas the solid lines display a two parameter fit. The
      crosses give the computed values of $B_c$ according to table~\ref{ferrodaten}.
      } \label{w_0-res}
\end{figure}

If the driving velocity coincides with this velocity, the system is in resonance and the amplitude exhibits its maximum. The phase velocity $v_p$ for different magnetic fields is experimentally obtained by fitting (\ref{amplitudeeq}) to the data. The results are shown in figure~\ref{w_0-res}. With increasing $B$ the velocity $v_p$ decreases and eventually becomes zero at a critical induction $B_{c,m}$.  At this induction the surface exhibits non-propagating undulations, which is a manifestation of the Rosensweig instability. 

Fitting the data with (\ref{wurzelfit}), where $\alpha$ is computed from the material parameters from table\,\ref{ferrodaten} and only $B_c$ is adjusted, yields the curves marked by dashed lines. Fitting both $B_c$ and $\alpha$ gives the solid lines. The values corresponding to the latter procedure are included as $B_{c,m}$ in table~\ref{ferrodaten}.

\section{Discussion and Conclusion}
Applying a novel type of magnetic traveling-stripe forcing with $k=k_c$ to the subcritical regime of the Rosensweig instability we measured the response of surface waves at different driving velocities $v$. For a driving at the phase velocity of free surface waves a resonance phenomenon is observed, which can quantitatively be described as that of a damped harmonic oscillator. The resonant velocity $v_p$ depends on the applied magnetic induction and decreases to zero when the critical induction $B_c$ is approached. The functional dependence of $v_p$ is essentially captured by the dispersion relation for an inviscid magnetic layer of infinite depth \cite{cowley1967}.

When comparing the values $B_c$ as computed from the material parameters and the fitted values $B_{c,m}$, the latter are shifted by 8\,\% and 9\,\% to lower inductions. This deviation is larger than the discrepancy of 3\,\% obtained from amplitude measurements in the supercritical regime of the Rosensweig instability \cite{gollwitzer2007}. It may partly be explained by the limited resolution in the immediate vicinity of $B_c$. More importantly the true value for $B_c$ can only be obtained in the limiting case for vanishing modulation amplitudes $\Delta B$, while $\Delta B$ in our case is approximately as large as the deviation $B_c-B_{c,m}$. In addition ansatz (\ref{wurzelfit}) neither takes into account the finite viscosity and layer depth \citeaffixed{lange2000}{see} nor a nonlinear magnetization law as utilized by \citeaffixed{gollwitzer2007,knieling2006}{see}. Further deviations may stem from the difficulty in producing an ideal rod lattice, leading to a scatter in the wavelength $\lambda$ by 4.3\,\% (rms). From figure~\ref{amplitude} we conclude that the thin ferrofluid EMG\,909 is more susceptible to these irregularities than the highly viscous fluid APG\,512a. This may as well be the reason for the difficulties with the one-parameter fit for this type of ferrofluid.

There are several possible applications of our spatiotemporal forcing. On the one hand, it can be used to create a volume flow \cite{kikura1990}.  A theoretical  estimate for this flow was recently provided by \citeasnoun{zimmermann2004} and is awaiting an experimental proof. This method of pumping is an alternative to those proposed by \citeasnoun{mao2005} and \citeasnoun{liu_patent}, the latter was realized by \citeasnoun{krauss2005}.

Further, the spatiotemporal forcing opens up new possibilities for the general study of the Rosensweig instability.
As an advantage, it may be used to fix the wavelength to preselected values. This is an important difference to a previous study by \citeasnoun{reimann2003}, where the critical scaling of a freely propagating wave has been measured, and a dependence $k(B)$ had to be taken into account. Utilizing a $k$ in the proper range may give an elegant access to the anomalous branch of the dispersion relation, which has only recently been studied by gravitational excitation in combination with different filling levels \cite{embs2007}. Moreover our magnetic forcing may be applied to the supercritical regime of the Rosensweig instability, where new resonances  between hexagonal, square, or stripe-like patterns and the traveling-stripe forcing are predicted \cite{ruediger2007}.

We thank Christopher Groh and Klaus Oetter for technical assistance with the setup. Financial support by \textit{Deutsche Forschungsgemeinschaft} under grant Ri 1054/1-4 is gratefully  acknowledged.

\section*{References}
\bibliographystyle{jphysicsB}

\end{document}